\documentclass[10pt]{iopart}
\usepackage{graphicx}
\usepackage{array}
\usepackage{url}
\usepackage{bm}
\usepackage{iopams}
\usepackage{tikz}

\renewcommand{\vec}[1]{\bm{#1}}

\begin{document}
\title{Improvements for drift-diffusion plasma fluid models with explicit time
  integration}

\author{Jannis Teunissen$^{1,2,3}$}

\address{$^1$Centrum Wiskunde \& Informatica (CWI), P.O. Box 94079, 1090 GB
  Amsterdam, The Netherlands}

\address{$^2$Centre for mathematical Plasma Astrophysics, Department of
  Mathematics, KU Leuven, Celestijnenlaan 200B, B-3001 Leuven, Belgium}

\address{$^3$State Key Laboratory of Electrical Insulation and Power Equipment,
  Xi'an Jiaotong University, Xi'an, China}

\ead{jannis@teunissen.net}

\begin{abstract}
  Drift-diffusion plasma fluid models are commonly used to simulate electric
  discharges. Such models can computationally be very efficient if they are
  combined with explicit time integration. This paper deals with two issues that
  often arise with such models. First, a high plasma conductivity can severely
  limit the time step. A fully explicit method to overcome this limitation is
  presented. This method is compared to the existing semi-implicit method, and
  it is shown to have several advantages. A second issue is specific to models
  with the local field approximation. Near strong density and electric field
  gradients, electrons can diffuse parallel to the field, and unphysically
  generate ionization. Existing and new approaches to correct this behavior are
  compared. Details on the implementation of the models and the various
  approaches are provided.
\end{abstract}

\maketitle
\ioptwocol

\section{Introduction}
\label{sec:intro}

Simulations of electric discharges are often performed with plasma fluid models
\cite{Gogolides_1992,Dujko_2013,Alves_2018}. Such models require that the mean
free path of electrons is small compared to characteristic length scales of the
discharge, so they typically become more accurate at higher pressures
(e.g.~$1 \, \textrm{bar}$). This paper specifically considers drift-diffusion
(DD) models with the local field approximation (LFA), which are referred to as
DD-LFA models. Such models are commonly used to simulate e.g., streamer
discharges~\cite{Luque_2012,Bagheri_2018,Marskar_2019a,Plewa_2018,Komuro_2018},
plasma jets~\cite{Boeuf_2013,Bourdon_2016} and dielectric barrier discharges
\cite{Stollenwerk_2006}.

Compared to particle simulations, fluid models are often much more
computationally efficient~\cite{Kim_2005}. Only a few densities have to be
evolved per grid cell, instead of tens or hundreds of particles. Updating these
densities in time is relatively cheap when an \emph{explicit} time
integrator is used, with which the state at time $t + \Delta t$ can
explicitly be constructed from the previous state(s). The numerical
implementation of a DD-LFA model is discussed in section \ref{sec:lfa-model}.

However, due to the coupling of charged species to the electric field, an
explicit time integrator leads to a restriction on the time step
$\Delta t \leq \varepsilon_0/\sigma$, where $\varepsilon_0$ is the permittivity of
vacuum and $\sigma$ the plasma conductivity. This restriction is discussed in
more detail in section \ref{sec:diel-relax-time}. A new explicit approach is
introduced, which avoids the time step restriction by limiting the conductivity
of the plasma. The new method is compared to the existing semi-implicit
method~\cite{Ventzek_1994,Lapenta_1995a,Hagelaar_2000a} in several test cases.

Near strong electric field and density gradients, the LFA loses (some of) its
validity. An unphysical effect that can occur is that electrons diffuse parallel
to the electric field into a high-field region, where they generate ionization.
This problem is discussed in detail in section \ref{sec:diffusion-grad}.
Existing and new approaches to correct this behavior are discussed and tested.

For simplicity, the test cases presented in this paper are one-dimensional.
However, the issues addressed are particularly relevant for 2D and 3D
simulations, which require both high computational efficiency (and thus explicit
schemes) and robustness, due to the sharp gradients and the geometrical
complexity that can occur.

\section{The DD-LFA model}
\label{sec:lfa-model}

In this section, a simple DD-LFA (drift-diffusion with local field
approximation) model is introduced. Detailed discussions of the validity of this
model, which are outside the scope of the present paper, can be found in
e.g.~\cite{Kim_2005,Grubert_2009,Becker_2013b,Markosyan_2015,Eichwald_2005,Li_2007,Drallos_1995}.
A brief summary is given below.

The accuracy of the LFA depends on the time scale of electron energy relaxation
(to the local conditions) compared to other time scales of interest. This energy
relaxation occurs through electron-neutral collisions, in particular inelastic
ones. In molecular gases at e.g.~$1 \, \textrm{bar}$, the LFA therefore works
better than in a noble gas at low pressure, if other conditions are kept the
same. However, even under favorable conditions, the LFA cannot capture certain
(non-local) effects, for example due to spatial density gradients or due to
spatial or temporal electric field gradients.

On the other hand, an advantage of the LFA is that only a single equation has to
be solved for electrons. Perhaps more important is that the electric field
strength is a relatively well-behaved parameter for determining transport
coefficients: it is non-negative, well-defined everywhere, and there is a direct
link to measured or computed transport coefficients in uniform fields.

\subsection{Model formulation}

The DD-LFA model considered here was chosen to be as simple as possible. Only
electrons and a single positive immobile ion species are included
\begin{eqnarray}
  \label{eq:fluid}
  \partial_t n_e &= -\nabla \cdot \vec{\Gamma} + S,\\
  \partial_t n_p &= S\nonumber,
\end{eqnarray}
where $n_e$ is the electron density, $n_p$ the positive ion density,
$\vec{\Gamma}$ the electron flux and $S$ the ionization source term. With the
drift-diffusion approximation, $\vec{\Gamma}$ is given by a drift and a
diffusion component
\begin{equation}
  \label{eq:flux-drift-diffusion}
  \vec{\Gamma} = \vec{\Gamma}^{\mathrm{drift}} + \vec{\Gamma}^{\mathrm{diff}}
  = - n_e \mu_e \vec{E} - D_e \nabla n_e,
\end{equation}
where $\mu_e$ is the electron mobility, $\vec{E}$ the electric field vector and
$D_e$ the electron diffusion coefficient. The electron-impact ionization term is
\begin{equation}
  \label{eq:source}
  S = \bar{\alpha} \mu_e |\vec{E}| n_e = \bar{\alpha} |\vec{\Gamma}^{\mathrm{drift}}|,
\end{equation}
where $\bar{\alpha}$ is the effective ionization coefficient (i.e., ionization
minus attachment), and $|\vec{E}|$ is the norm of the electric field. With the
local field approximation, the electron velocity distribution is assumed to be
relaxed to the local electric field, so that $\mu_e$, $D_e$ and $\bar{\alpha}$
are functions of $|\vec{E}|$.

For general (multidimensional) simulations, the electric field is computed from
the electrostatic potential $\phi$ as $\vec{E} = -\nabla \phi$, and $\phi$ is
obtained by solving a Poisson equation. In simple cases without dielectrics or
electrodes, this Poisson equation looks as follows
\begin{equation}
  \label{eq:poisson}
  \nabla^2 \phi = -\rho/\varepsilon_0,
\end{equation}
where $\varepsilon_0$ is the permittivity of vacuum, $\rho = (n_p - n_e)e$ the
charge density and $e$ the elementary charge. For the 1D simulations considered
here, the situation simplifies, as described below.

\subsection{Finite volume implementation}

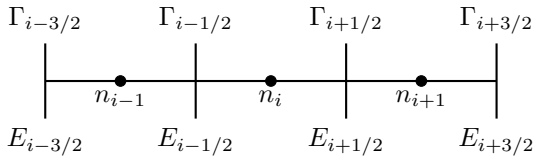
\begin{figure}
  \centering
  \begin{tikzpicture}
    % Horizontal line
    \draw[thick] (0,0) -- (6,0);
    % Vertical lines
    \foreach \x in {0, 2, 4, 6}
    \draw[thick] (\x,-0.5) -- (\x,0.5);
    % Cell centers
    \foreach \x in {1, 3, 5}
    \filldraw (\x,0) circle (2pt);
    % Densities
    \node[anchor=north] at (1,0) {$n_{i-1}$};
    \node[anchor=north] at (3,0) {$n_{i}$};
    \node[anchor=north] at (5,0) {$n_{i+1}$};
    % Fluxes
    \node[anchor=south] at (0,0.5) {$\Gamma_{i-3/2}$};
    \node[anchor=south] at (2,0.5) {$\Gamma_{i-1/2}$};
    \node[anchor=south] at (4,0.5) {$\Gamma_{i+1/2}$};
    \node[anchor=south] at (6,0.5) {$\Gamma_{i+3/2}$};
    % Fields
    \node[anchor=north] at (0,-0.5) {$E_{i-3/2}$};
    \node[anchor=north] at (2,-0.5) {$E_{i-1/2}$};
    \node[anchor=north] at (4,-0.5) {$E_{i+1/2}$};
    \node[anchor=north] at (6,-0.5) {$E_{i+3/2}$};
  \end{tikzpicture}
  \caption{Illustration of a finite-volume discretization in 1D. Densities $n_i$
    are stored at cell centers. Fluxes $\Gamma_{i+1/2}$ and electric fields
    $E_{i+1/2}$ are located at cell faces.}
  \label{fig:fv-discretization}
\end{figure}

Fluid simulations of electric discharges are often performed with finite volume
(FV) methods, see e.g.~\cite{LeVeque_2002,Toro_2009}. With a FV method, fluxes
are computed at the faces of a cell, and the volume-averaged densities inside a
cell are updated with these fluxes, see figure \ref{fig:fv-discretization}.

The flux is here computed using a slope limiter, as described in
\cite{Montijn_2006,Teunissen_2017}. The idea behind slope limiters is to
interpolate cell-centered densities to the cell faces (where the flux has to be
computed) in such a way that numerical errors do not grow in time. The scheme's
implementation in 1D is as follows. If $v = -\mu_e E_{i+1/2}$ is the drift
velocity at cell face $i+1/2$, then $\Gamma_{i+1/2}^\mathrm{drift}$ is computed
as
\begin{equation*}
  \label{eq:koren-neg}
  \Gamma_{i+1/2}^\mathrm{drift} = \cases{
    v \left[n_{i+1} - \psi(1/r_{i+1}) (n_{i+1} - n_{i})\right]&$v < 0$\\
    v \left[n_{i} + \psi(r_i) (n_{i+1} - n_{i})\right]&$v \geq 0$}
\end{equation*}
where $r_i = (n_{i} - n_{i-1})/(n_{i+1} - n_{i})$ and $\psi(x)$ is the limiter
function. For brevity of notation, the electron density in cell $i$ is here (and
later in the paper) indicated by $n_i$. Note that if $\psi(x) = 0$, the flux is
given by the first-order upwind method. As
in~\cite{Montijn_2006,Teunissen_2017}, the Koren~\cite{Koren_1993} limiter is
used, given by
\begin{equation*}
  \psi(x) = \max\left(0, \min(1, (2+x)/6, x)\right).
\end{equation*}
The diffusive flux is computed as
\begin{equation*}
  \Gamma_{i+1/2}^\mathrm{diff} = -D_e (n_{i+1} - n_{i})/\Delta x,
\end{equation*}
where $\Delta x$ is the grid spacing. Note that with the LFA, $\mu_e$ and $D_e$
depend on $|\vec{E}|$ at the cell face. In 1D, one can simply take the absolute
value, but in multiple dimensions this requires some type of interpolation for
the extra components, see e.g.~\cite{Teunissen_2017}.

Unless specified otherwise, source terms are evaluated at the cell centers.
Equation (\ref{eq:source}) is then implemented as
\begin{equation}
  \label{eq:source-cc}
  S_i = \bar{\alpha} \mu_e |E_i| n_i,
\end{equation}
where $\bar{\alpha}$ and $\mu_e$ depend on the electric field strength at the
cell center $|E_i|$, which is here computed as
$|E_i| = |E_{i-1/2} + E_{i+1/2}|/2$. As will be shown in sections
\ref{sec:diel-relax-time} and \ref{sec:diffusion-grad}, it is sometimes
beneficial to evaluate the source term at cell faces using the electron flux. On
a grid with square cells (e.g., $\Delta x = \Delta y$), equation
(\ref{eq:source}) can then be implemented as
\begin{equation}
  \label{eq:source-gamma}
  S_i = 1/N \sum \bar{\alpha}_{\mathrm{face}} |\Gamma_{\mathrm{face}}^\mathrm{drift}|,
\end{equation}
where the sum runs over all $N$ cell faces, $\bar{\alpha}_{\mathrm{face}}$
depends on the electric field at the cell face, and
$\Gamma_{\mathrm{face}}^\mathrm{drift}$ is the drift flux through the cell face.

Time integration is here performed with the explicit trapezoidal rule, as
in~\cite{Montijn_2006,Teunissen_2017}. If equation (\ref{eq:fluid}) is written
as $\mathbf{y}'(t) = \mathbf{f}(\mathbf{y})$, then this scheme is given by
\begin{eqnarray*}
  \tilde{\mathbf{y}}_{t+1} &= \mathbf{y}_t + \Delta t \, \mathbf{f}(\mathbf{y}_t)\\
  \mathbf{y}_{t+1} &= \mathbf{y}_t + \frac{\Delta t}{2}
                     \left[\mathbf{f}(\mathbf{y}_t) +
                     \mathbf{f}(\tilde{\mathbf{y}}_{t+1})\right].
\end{eqnarray*}
Note that this is an \emph{explicit} scheme, i.e., that $\mathbf{y}_{t+1}$ can
explicitly be computed from $\mathbf{y}_{t}$. Additionally, the classic
fourth-order Runge-Kutta time integrator (RK4) is used for some of the
convergence tests in section \ref{sec:diel-relax-time}. For most of the tests performed here, the time step $\Delta t$ is fixed. In other cases, it is determined as
\begin{equation}
  \label{eq:cfl}
  \Delta t = 0.9 \times \min\left(0.5 \, \tau_\mathrm{CFL},
    \frac{1}{1/\tau_\mathrm{CFL}+1/\tau_\mathrm{D}}\right),
\end{equation}
where $\tau_\mathrm{CFL}$ and $\tau_\mathrm{D}$ are given by the minimum values
of $\Delta x / v$ and $\Delta x^2/(2 D_e)$, respectively. Another time step
restriction related to the plasma conductivity is sometimes required, see
section \ref{sec:diel-relax-time} and specifically equation (\ref{eq:dt-req}).

A new electric field is computed after each (sub)step of the time integrator. In
1D, this is done by starting from a guess for the electric field at the boundary
of the domain. The electric fields at all other cell faces can then be obtained
from
\begin{equation*}
  E_{i+1/2} = E_{i-1/2} + \Delta x \, \rho_i/\varepsilon_0,
\end{equation*}
after which the total voltage difference over the domain is determined. Finally,
a constant background field is added to ensure the voltage difference becomes
the desired applied voltage.

\section{The dielectric relaxation time}
\label{sec:diel-relax-time}

In a plasma, the movement of charged species is tightly coupled to the electric
field. In general, charges move to screen electric fields that are present
within the plasma. A characteristic time scale for this screening is the
dielectric relaxation time, also known as the Maxwell time~\cite{Barnes_1987,Teunissen_2014}
\begin{equation}
  \label{eq:tau-sigma}
  \tau = \varepsilon_0/\sigma,
\end{equation}
where $\sigma$ is the plasma conductivity. For the discharges considered here,
the conductivity is given by
\begin{equation*}
  \sigma = e \sum_k \mu_k n_k,
\end{equation*}
where $\mu_k$ and $n_k$ are the species mobilities and densities. Typically, the
contribution of the electron mobility $\mu_e$ dominates, so that
$\sigma \approx e \mu_e n_e$. In section \ref{sec:tau-deriv}, it is shown that
for plasma models with explicit time integration, a time step restriction
$\Delta t \leq \varepsilon_0/\sigma$ is required.

\subsection{Derivation}
\label{sec:tau-deriv}

Equation (\ref{eq:tau-sigma}) is derived along the lines of
\cite{Teunissen_2014}. First, recall Maxwell-Amp{\`e}re's equation
\begin{equation}
  \label{eq:ampere}
  \nabla \times \vec{B} = \mu_0 (\vec{J} + \varepsilon_0 \partial_t \vec{E}),
\end{equation}
where $\vec{B}$ is the magnetic field, $\mu_0$ the permeability of free space
and $\vec{J}$ the electric current density. Since the divergence of a curl is
zero, it follows that
\begin{equation}
  \label{eq:div-zero}
  \nabla \cdot (\vec{J} + \varepsilon_0 \partial_t \vec{E}) = 0.
\end{equation}
In the electrostatic approximation with $\vec{B} = 0$ it can be assumed that $\vec{J}$ and $\vec{E}$ are (anti-)parallel, since the drift flux is generally (anti-)parallel to $\vec{E}$ and since the same typically holds for the diffusive fluxes. The divergence can then be transformed into a single spatial
derivative, and the generic solution of equation (\ref{eq:div-zero}) is
\begin{equation}
  \label{eq:dtE-J}
  \partial_t E = -J / \varepsilon_0 + C,
\end{equation}
where $C$ depends on changes elsewhere in the system\footnote{In 1D, $C = 0$ as
  long as the boundary conditions do not change in time.}. Plugging in
$J = \sigma E + J^\mathrm{diff}$, where $J^\mathrm{diff}$ are diffusive terms
that do not depend on $E$, gives
\begin{equation}
  \label{eq:dtE-E}
  \partial_t E = -\frac{\sigma}{\varepsilon_0} E - J^\mathrm{diff}/\varepsilon_0 + C.
\end{equation}
When systems with the dynamics of equation (\ref{eq:dtE-E}) are numerically
integrated with an explicit method, it is typically required that
\begin{equation}
  \label{eq:dt-req}
  \Delta t \leq \varepsilon_0/\sigma,
\end{equation}
otherwise errors grow in time. A physical interpretation of this restriction is
that for larger time steps the electric field can reverse, and even be amplified
in the reverse direction, which leads to growing errors.

% In a non-planar geometry, the above derivation still holds since $\vec{E}$ and
% $\vec{J}$ are typically parallel. The coefficient $C$ of equation
% (\ref{eq:dtE-J}) then depends on the displacement of charges elsewhere, but this
% does not affect the stability criterion

% We remark that electric screening happens most quickly in planar geometries, in
% which all of the current $\vec{J}$ contributes to $\partial_t \vec{E}$.

\subsection{Semi-implicit approach}
\label{sec:semi-impl}

The time step restriction of equation (\ref{eq:dt-req}) can be avoided by
solving the plasma fluid equations fully implicitly in time. However, obtaining
such implicit solutions is usually computationally expensive, in particular in
2D and 3D, which is in part due to the non-linear coupling with the electric
field. For this reason, \emph{semi-implicit}
discretizations~\cite{Ventzek_1994,Lapenta_1995a,Hagelaar_2000a} have frequently
been used. The idea is to first predict the electric field at $t+\Delta t$ by
solving a Poisson equation
\begin{equation}
  \label{eq:semi-impl-pois}
  -\nabla^2 \tilde{\phi} = \nabla \cdot \vec{\tilde{E}} =
  \frac{1}{\varepsilon_0} \sum_k q_k \tilde{n}_k,
\end{equation}
where tildes ($\tilde{\;}$) indicate estimates at $t+\Delta t$, $\phi$ is the electric
potential, $q_k$ the charge of species $k$ and $\tilde{n}_k$ its density. Since
source terms create no net charge, densities can predicted using only the flux
$\Gamma$, see equation (\ref{eq:flux-drift-diffusion}):
\begin{equation}
  \label{eq:semi-impl-estimate}
  \tilde{n}_k = n_k^t + \Delta t \, \nabla \cdot \vec{\Gamma}_k(n_k^t, \mu_k^t, D_k^t, \vec{\tilde{E}}),
\end{equation}
where the superscript $t$ indicates quantities known at the present time.
Combining equations (\ref{eq:semi-impl-estimate}), (\ref{eq:semi-impl-pois}) and
(\ref{eq:flux-drift-diffusion}) leads to a variable-coefficient elliptic PDE,
which can be solved to obtain $\tilde{\phi}$ and $\vec{\tilde{E}}$. Afterwards,
the fluid equations can be solved using $\vec{\tilde{E}}$. Such a semi-implicit
approach has the following properties:
\begin{itemize}
  \item The time step restriction of equation (\ref{eq:dt-req}) is avoided.
  \item The scheme is first order accurate in time.
  \item The resulting variable-coefficient elliptic PDE can be solved quite
  efficiently, but the cost is typically higher than for constant-coefficient
  cases such as equation (\ref{eq:poisson}).
\end{itemize}

\subsection{Current-limited approach}
\label{sec:limit-sigma}

Instead of solving equations implicitly or imposing a restriction on the time
step $\Delta t$, another option is to limit the conductivity to allow for a
larger time step. Equation (\ref{eq:dt-req}) then becomes
\begin{equation}
  \label{eq:sigma-req}
  \sigma \leq \varepsilon_0/\Delta t.
\end{equation}
If both sides are multiplied with $|E_j|$, where $E_j$ is the $j$\textsuperscript{th} component
of the electric field, and it is assumed that $\sigma \approx e \mu_e n_e$,
equation (\ref{eq:sigma-req}) can be rewritten as
\begin{equation}
  \label{eq:sigma-req-2}
  \mu_e |E_j| n_e \leq \varepsilon_0 |E_j| /(e \Delta t).
\end{equation}
The left-hand side of equation (\ref{eq:sigma-req-2}) is equal to
$|\Gamma_j^{\mathrm{drift}}|$, where $\Gamma_j^{\mathrm{drift}}$ is the $j$\textsuperscript{th}
component of the electron drift flux, see equation
(\ref{eq:flux-drift-diffusion}). Instead of limiting only the drift component,
it is often preferable to limit the \emph{total} electron flux
$|\Gamma_j| = |\Gamma_j^{\mathrm{drift}} + \Gamma_j^{\mathrm{diff}}|$, because
diffusive fluxes can also be large near strong density gradients. This is here
done according to the following expression
\begin{equation}
  \label{eq:mu-req}
  |\Gamma_j| \leq \varepsilon_0 E^*/(e \Delta t),
\end{equation}
where $E^*$ should also take diffusion into account. If diffusion is dominant,
electron drift and diffusive fluxes will quickly balance each other, and a
reasonable choice for $E^*$ is the electric field at which this occurs. Solving
equation (\ref{eq:flux-drift-diffusion}) for $|\Gamma_j| = 0$ gives
$|E_j| = D_e |\partial_j n_e| / (n_e \mu_e)$, where $\partial_j$ is the
derivative in direction $i$. Conversely, if the drift flux is dominant, $E^*$
should be set to $|E_j|$, as was done above. As a general expression for $E^*$,
the maximum of these two cases is used
\begin{equation}
  \label{eq:E-star}
  E^* = \max\left(|E_j|, \frac{D_e |\partial_j n_e|}{\mu_e n_e}\right).
\end{equation}
A robust way to compute $|\partial_j n_e|/n_e$ at the face between cells $i$ and
$i+1$, where the flux is defined in finite volume schemes, is
\begin{equation}
  \label{eq:balance-comp}
  \frac{|\partial_j n_e|}{n_e} \approx \frac{|n_{i+1} - n_i|/\Delta x}{\max(n_i, n_{i+1},\epsilon)},
\end{equation}
where $\Delta x$ is the grid spacing and $\epsilon$ is a small number to avoid
division by zero. Note that the other quantities in equation (\ref{eq:E-star})
can simply be reused from the normal flux computation.

\subsubsection{Implementation}
\label{sec:current-lim-impl}

For convenience, the implementation of the suggested scheme in a discharge model
is summarized below:
\begin{enumerate}
  \item \label{enum:step1} Determine a time step $\Delta t$ without taking the
  dielectric relaxation time $\tau = \varepsilon_0/\sigma$ into account, for
  example as in equation (\ref{eq:cfl}).
  \item Compute all the components of the electron flux $\Gamma_j$ at the cell
  faces between the grid cells, for example as described in section
  \ref{eq:koren-neg}.
  \item Compute the values of $E^*$ at the cell faces, using equations
  (\ref{eq:E-star}) and (\ref{eq:balance-comp}).
  \item Limit the components of the electron flux $\Gamma_j$ so that equation
  (\ref{eq:mu-req}) holds. In other words, if
  $|\Gamma_j| > \varepsilon_0 E^*/(e \Delta t)$, set
  $\Gamma_j = \mathrm{sgn}(\Gamma_j) \, \varepsilon_0 E^*/(e \Delta t)$, where
  $\mathrm{sgn}(x)$ is the sign function.
\end{enumerate}
Sometimes, it can be important to resolve electric screening accurately in time,
for example when a highly conductive plasma region is forming. In such cases,
the $\Delta t$ in step~(\ref{enum:step1}) can temporarily be reduced.

\subsubsection{Discussion}
\label{sec:current-lim-discuss}

The current-limited approach limits the transported charge between two adjacent
grid cells within a single time step. Per unit area, the transported charge is
$\delta_\sigma = e \, \Delta t \, |\Gamma_j|$, where $\Gamma_j$ is the electron
flux in direction $j$. From equation (\ref{eq:mu-req}), it follows that
$\delta_\sigma \leq \varepsilon_0 E^*$, which corresponds to a change in
local electric field satisfying $|\delta_E| \leq E^*$. Two cases can be distinguished for
$E^*$, see equation (\ref{eq:E-star}). When $E^* = |E_j|$, it follows that the
previous field is at most completely screened. When
$E^* = D_e |\partial_j n_e|/(\mu_e n_e)$, the change in field is limited by the
field that would balance drift and diffusive fluxes.

The current-limited scheme can be applied to drift-diffusion models in general,
regardless of whether they use the local field approximation. Advantages of the
scheme are that:
\begin{itemize}
  \item The time step restriction of equation (\ref{eq:dt-req}) is avoided.
  \item For time steps fulfilling equation (\ref{eq:dt-req}) the original
  behavior of a fluid model is restored.
  \item It requires only a minor modification of the flux computation.
  \item The computational cost is low, as the scheme is fully explicit.
\end{itemize}

\subsection{Comparison of current-limited and semi-implicit approach}
\label{sec:drt-test-one}

The current-limited approach is now compared with the semi-implicit method in a
simplified 1D test case. Equations
(\ref{eq:fluid}--\ref{eq:flux-drift-diffusion}) are solved with the source term
set to zero and constant transport coefficients
$\mu_e = 0.03 \, \mathrm{m}^2/(\mathrm{V \, s})$ and
$D_e = 0.1 \, \mathrm{m}^2/\mathrm{s}$. A computational domain of
$10 \, \textrm{mm}$ is used, with a grid spacing
$\Delta x = 20 \, \mu\textrm{m}$. The right boundary is grounded, and a voltage
of $10 \, \textrm{kV}$ is applied at the left boundary. The initial densities
$n_e$ and $n_p$ are $10^{20} \, \textrm{m}^{-3}$ for
$4 \, \textrm{mm} \leq x \leq 6 \, \mathrm{mm}$, and zero elsewhere. With these
parameters, the dielectric relaxation time of equation (\ref{eq:dt-req}) becomes
$\tau \approx 18.4 \, \mathrm{ps}$.

\begin{figure}
  \centering
  \includegraphics{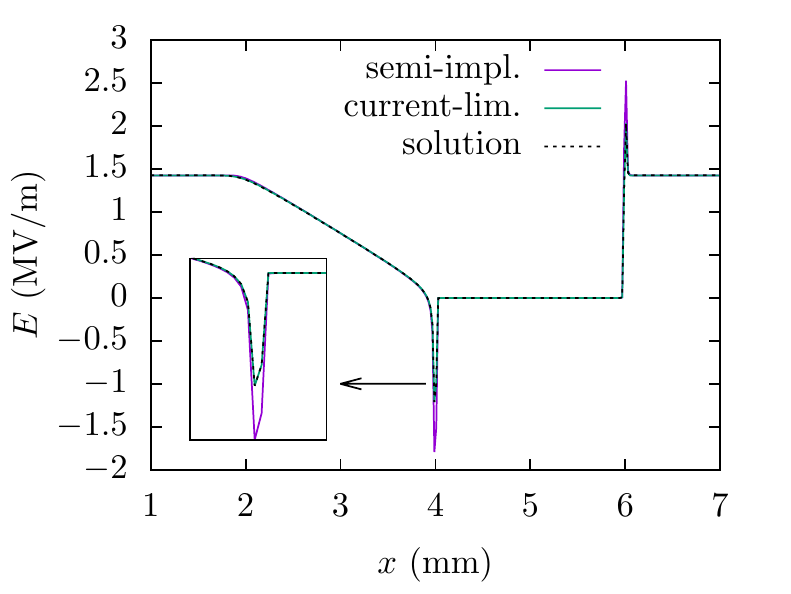}
  \includegraphics{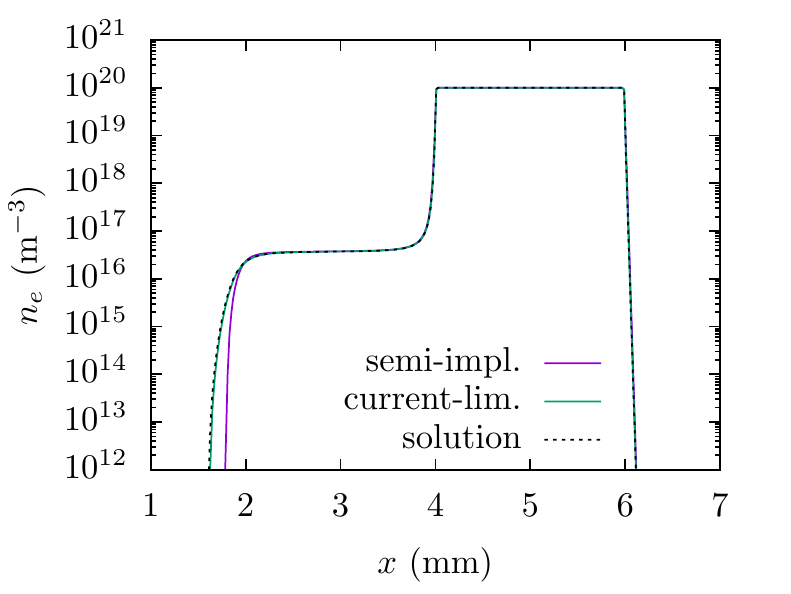}
  \caption{Comparison of the semi-implicit approach with the new current-limited
    approach. Shown are the electric field (top) and the electron density
    (bottom) at $t = 50 \, \textrm{ns}$ for the test case described in section
    \ref{sec:drt-test-one}. For both approaches, a time step
    $\Delta t = 80 \, \textrm{ps}$ is used. A reference solution with a small
    time step is also shown. With the semi-implicit approach, the electric field
    has larger peaks at the boundary of the high-density region.}
  \label{fig:drt-test-one}
\end{figure}

Figure \ref{fig:drt-test-one} shows the electric field and the electron density
at $t = 50 \, \textrm{ns}$ for the current-limited and semi-implicit approach
using a time step $\Delta t = 80 \, \textrm{ps}$. For comparison, a reference
solution computed with the fourth-order RK4 time integrator and a small time
step $\Delta t = 0.1 \, \textrm{ps}$ is also shown. This reference solution was
computed on the same grid, without special treatment for the dielectric
relaxation time. The current-limited approach agrees very well with the
reference solution, whereas the semi-implicit method predicts larger peaks in
the electric field at the boundaries of the initially ionized area. The left
side of the electron density profile also has a visibly different shape with the
semi-implicit approach.

\begin{figure}
  \centering
  \includegraphics{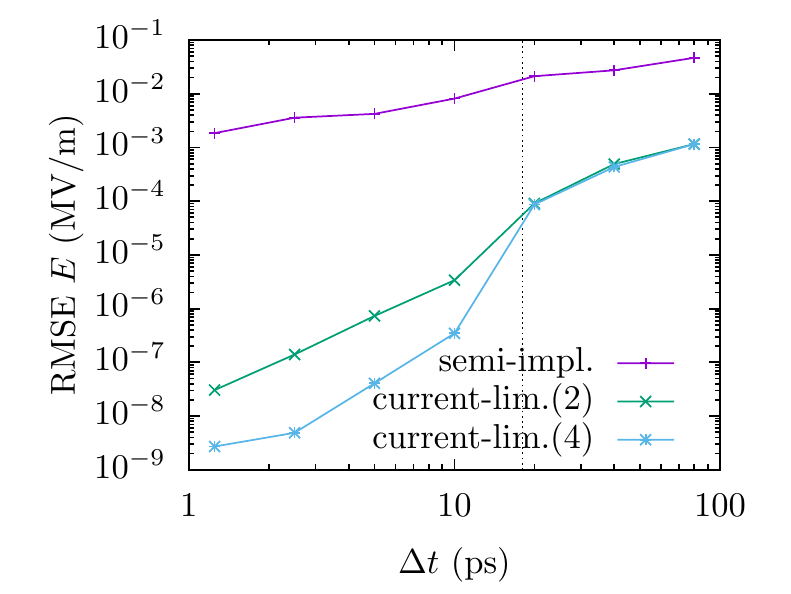}%
  \caption{A comparison of the convergence behavior of the semi-implicit and
    current-limited approaches for the test case described in
    section~\ref{sec:drt-test-one}. Shown is the root-mean-square error
    (compared to the reference solution) in the electric field at
    $t = 50 \, \textrm{ns}$ versus the time step. For the current-limited
    approach, results are shown with a second-order and a fourth-order accurate
    time integrator, indicated by (2) and (4), respectively. The vertical dashed
    line indicates the dielectric relaxation time.}
  \label{fig:drt-test-one-conv}
\end{figure}

Figure \ref{fig:drt-test-one-conv} shows the root-mean-square error in the
electric field (compared to the reference solution) versus the time step. When
the time step is reduced, both the current-limited and the semi-implicit
approach appear to converge to the reference solution, although with quite
different convergence rates. The semi-implicit method shows roughly first-order
convergence. Errors are significantly larger than with the current-limited
approach, also for time steps larger than the dielectric relaxation time $\tau$,
indicated by the vertical dashed line.

For the current-limited method, the order of convergence depends on the time
integrator for sufficiently small time steps. Two cases are shown: one with the
second-order time integrator described in section \ref{eq:koren-neg}, and one
with a fourth-order accurate time integrator (RK4). For the latter method, the
error saturates for the smallest time steps, which is probably due to numerical
round-off errors. For time steps larger than $\tau$, the second and fourth order
scheme behave similarly, which indicates that the limiting procedure is the main
source of errors.

\subsection{Test in nitrogen}
\label{sec:drt-test-two}

In this section, the current-limited approach is used in a more realistic test
case in nitrogen at $1 \, \textrm{bar}$ and $300 \, \textrm{K}$. Transport
coefficients and the ionization source term are computed with
Bolsig+~\cite{Hagelaar_2005} as function of the applied electric field, using
Phelps' cross sections~\cite{Phelps_1985}. The same initial condition and grid
are used as in section \ref{sec:drt-test-one}, but with an applied voltage of
$40 \, \textrm{kV}$. After electric screening has taken place in the initial
plasma region, the dielectric relaxation time is about
$\tau \approx 3 \, \mathrm{ps}$. This value depends on the highest tabulated
electron mobility (for the lowest electric field), which was here
$\mu_e \approx 0.186 \, \mathrm{m}^2/(\mathrm{V s})$.

\begin{figure}
  \centering
  \includegraphics{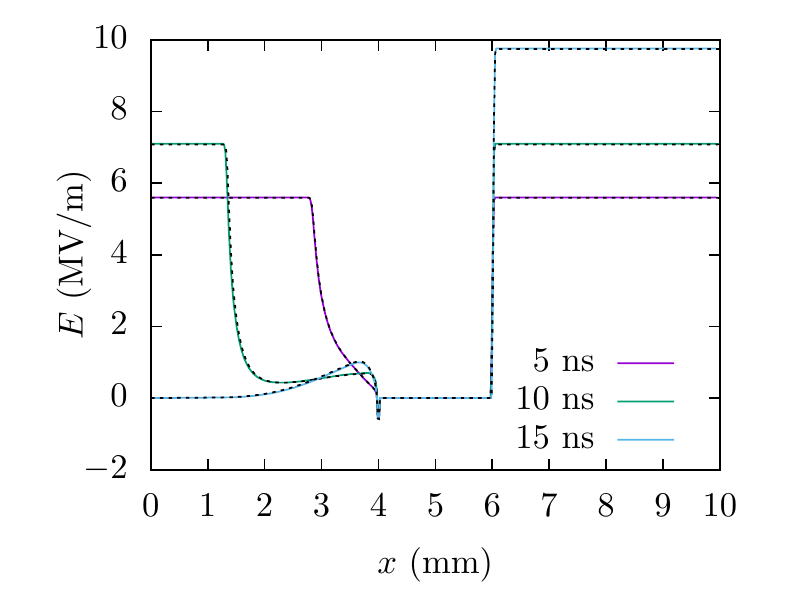}
  \includegraphics{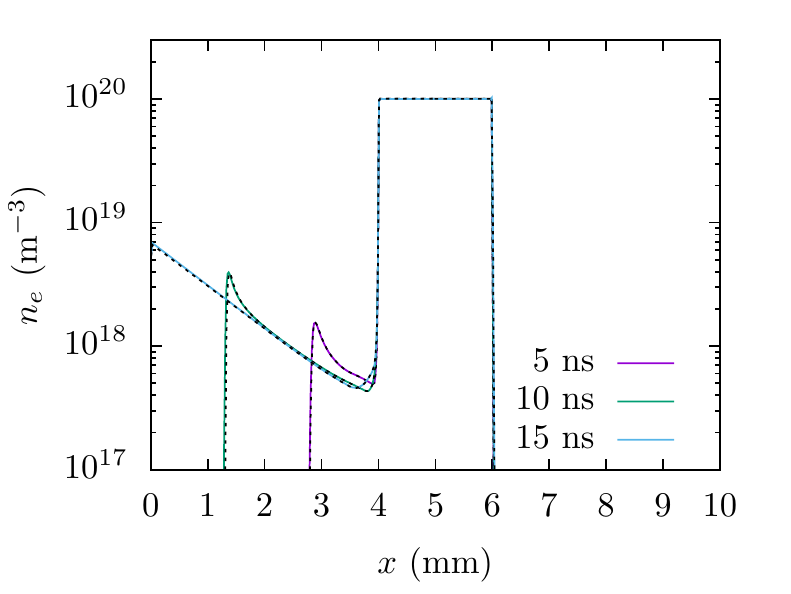}
  \caption{Results of the current-limited approach (solid lines) for a test case
    in nitrogen, see section \ref{sec:drt-test-two}. Shown are the electric field
    (top) and the electron density (bottom) at several times. The dashed lines
    indicate a reference solution with a small time step
    $\Delta t = 20 \, \mathrm{ps}$.}
  \label{fig:drt-test-n2}
\end{figure}

\begin{figure}
  \centering
  \includegraphics{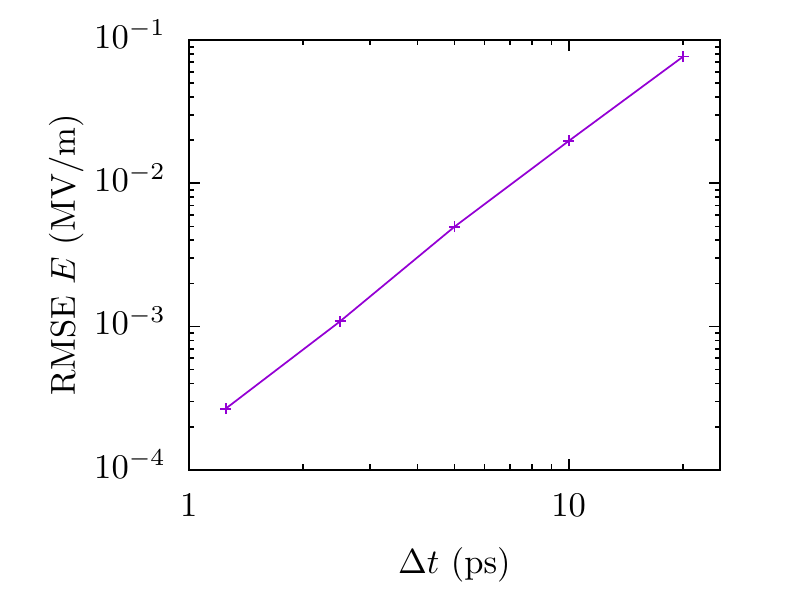}
  \caption{The convergence behavior of the current-limited approach for a test
    case in nitrogen, see section~\ref{sec:drt-test-two}. Shown is the
    root-mean-square error (compared to the reference solution) in the electric
    field at $t = 10 \, \textrm{ns}$ versus the time step. The line indicates
    that second order convergence is obtained.}
  \label{fig:drt-test-two-conv}
\end{figure}

Figure \ref{fig:drt-test-n2} shows the electron density and electric field in
the system at several times. The solid lines show results with the
current-limited approach and a fixed time step $\Delta t = 20 \, \mathrm{ps}$.
The dashed lines show a reference solution, computed on the same numerical grid
but with a small time step $\Delta t = 0.1 \, \mathrm{ps}$ and a fourth-order
time integrator (RK4). Visually, the current-limited results are almost
identical to the reference solution, indicating that the approach also works
well in more realistic test cases. The convergence to the reference solution is
shown in figure \ref{fig:drt-test-two-conv}. Second order convergence is
obtained, which indicates that the current-limited approach does not affect the convergence of the time integration scheme.

\subsection{Relevance for discharge simulations}
\label{sec:drt-relevance}

Avoiding the time step restriction due to the dielectric relaxation time $\tau$
is beneficial when $\tau$ becomes smaller than other time step constraints, in particular when $\tau < \tau_\mathrm{CFL}$, see equation (\ref{eq:cfl}). This can for example happen when:
\begin{itemize}
  \item High-density low-pressure discharges are simulated, which have a high conductivity, see e.g.~\cite{Ventzek_1994,Lapenta_1995a}.
  \item A localized high-density and thus high-conductivity region is present,
  for example near an electrode or a dielectric material, see e.g.
 ~\cite{Teunissen_2016,Babaeva_2016}, or when streamer discharges stagnate
 ~\cite{Pancheshnyi_2004}.
  \item The evolution of a conductive plasma is studied after the voltage has
  been turned off.
\end{itemize}

\section{Unphysical diffusion}
\label{sec:diffusion-grad}

The diffusive term in the electron flux of equation
(\ref{eq:flux-drift-diffusion}) can sometimes lead to unphysical behavior, in
particular when the diffusive flux has a component parallel to $\vec{E}$. This
can for example happen when there is an electrically screened region with a high
electron density, bordered by a region without electrons where the electric
field is above breakdown. In reality, electrons should quickly lose energy as
they diffuse parallel to the electric field, so that they are effectively
confined. However, a fluid model with the LFA does not capture these dynamics,
and allows electrons to diffuse into the high-field region\footnote{Moreover, a
  strong parallel electric field will typically give a higher diffusion
  coefficient with the LFA.}. These electrons then start generating
electron-impact ionization, since their ionization rate only depends on the
local electric field.

This unphysical effect can for example occur when simulating streamer discharges
close to dielectric surfaces~\cite{Babaeva_2016,Soloviev_2009,Soloviev_2017}.
The author has also encountered it in 3D simulations of branching streamer
discharges~\cite{Teunissen_2017}, in which some of the smaller branches stop to
grow~\cite{Pancheshnyi_2004}.

Below, the unphysical behavior is first demonstrated in a test case. Several
approaches to prevent the unphysical behavior are then discussed, and these
approaches are compared in section \ref{sec:tests-diff-schemes}.

\subsection{Test case}
\label{sec:diff-test-case}

The unphysical behavior due to parallel diffusion is present in the test case of
section \ref{sec:drt-test-two}, but it is not yet visible at
$t = 15 \, \textrm{ns}$ in figure \ref{fig:drt-test-n2}. To demonstrate the
unphysical effect more clearly, simulations are here performed in a higher
background field. A computational domain of $4 \, \textrm{mm}$ is used with a
voltage difference of $40 \, \textrm{kV}$, so that the background field of
$10 \, \textrm{MV/m}$ points in the $+x$-direction. A grid spacing of
$\Delta x = 4 \, \mu\mathrm{m}$ is used. Initially,
$n_e = n_p = 10^{20} \, \mathrm{m}^{-3}$ in the left half of the domain, and
$n_e = n_p = 0$ in the right half of the domain. As in section
\ref{sec:drt-test-two}, the simulations are performed in nitrogen at
$1 \, \textrm{bar}$ and $300 \, \textrm{K}$.

\begin{figure}
  \centering
  \includegraphics[width=0.5\textwidth]{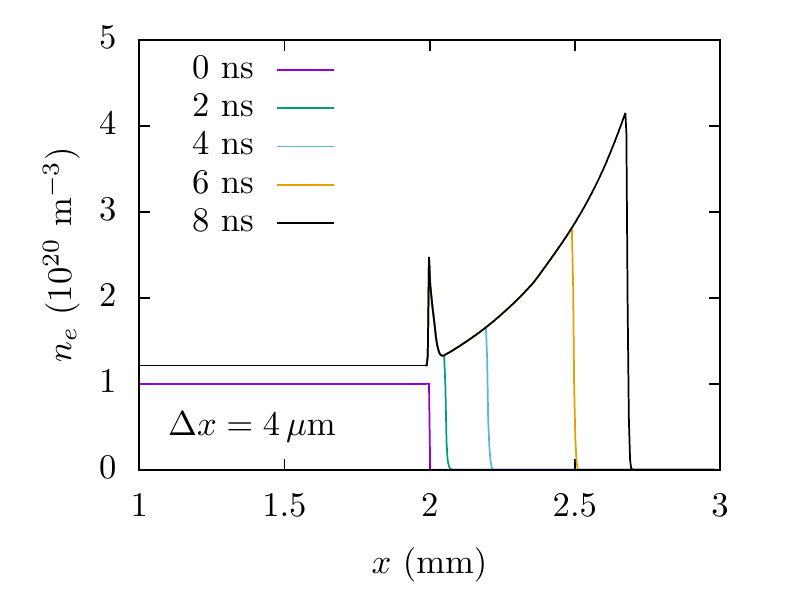}
  \includegraphics[width=0.5\textwidth]{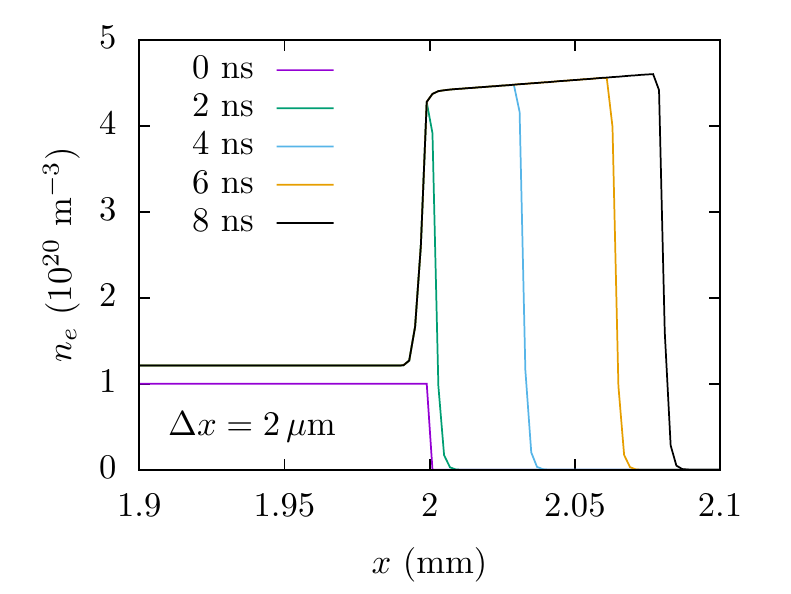}
  \caption{Illustration of the unphysical growth of an ionized region parallel
    to the electric field, which here points to the right. In the standard
    DD-LFA model, electrons can diffuse into the high-field region on the right,
    where they generate impact ionization. The grid spacing for the top and
    bottom figure are $4 \, \mu\textrm{m}$ and $2 \, \mu\textrm{m}$,
    respectively.}
  \label{fig:diffusion-ex}
\end{figure}

The evolution of the electron density is shown in figure \ref{fig:diffusion-ex},
which also includes results with half the grid spacing
($\Delta x = 2 \, \mu\textrm{m}$). In both cases, electrons diffuse to the right
of the original plasma boundary. When they enter the high-field region, they
generate ionization, extending the plasma to the right. Using a finer grid
spacing slows down the unphysical growth of the plasma, but does not prevent it.
Not shown here is the electric field, which is quickly screened in the left half
of the domain~\cite{Teunissen_2014}, after which it doubles in the right half of
the domain. During the initial electric screening, the degree of ionization in
the left half of the domain slightly increases.

\subsection{Source term factor ($f_\epsilon$ scheme)}
The LFA typically becomes less accurate when there are strong density and/or
field gradients~\cite{Naidis_1997,Li_2010a}. Soloviev and
Krivtsov~\cite{Soloviev_2009} derived a correction factor $f_\epsilon$ for the
impact ionization term from the electron energy equation, which is here referred
to as the $f_\epsilon$ scheme. This factor can be written as
\begin{equation}
  \label{eq:soloviev}
  f_\epsilon = 1 - \frac{\hat{E} \cdot \vec{\Gamma}^{\mathrm{diff}}}{|\Gamma^{\mathrm{drift}}|}
  = 1 + \frac{\hat{E} \cdot (D_e \nabla n_e)}{\mu_e n_e |E|},
\end{equation}
where $\hat{E}$ is the electric field unit vector. Near a strong density
gradient such as shown in figure \ref{fig:diffusion-ex}, $f_\epsilon$ will go to zero, so
that no unphysical ionization will take place. This happens because the
advective and the diffusive flux balance each other, with the diffusive flux
occurring parallel to $\vec{E}$. Note that equation (\ref{eq:soloviev}) also
gives reasonable results when the drift and diffusive fluxes do not completely
cancel. For example, if the diffusive flux balances half the drift flux
($\vec{\Gamma}^{\mathrm{diff}} = -\vec{\Gamma}^{\mathrm{drift}}/2$), the result
is $f_\epsilon = 1/2$.

A downside of using equation (\ref{eq:soloviev}) is that the original model
formulation is modified, also in regions where there are no unphysical effects
due to diffusion. Another issue is how to compute $f_\epsilon$ numerically. If
the right-most expression of equation (\ref{eq:soloviev}) is used, there are two
problems. First, quantities such as $n_e$, $\vec{E}$ and $\nabla n_e$ are not
all defined at the same location in a numerical grid cell, so that some type of
interpolation or averaging is required. Second, since both $n_e$ and $E$ can be
small (or even zero), the division can be problematic. However, $f_\epsilon$ can
robustly be computed at a cell face if the electron flux is used. In 1D,
equation (\ref{eq:soloviev}) can then be implemented as
\begin{equation}
  \label{eq:soloviev-scheme}
  f_\epsilon = 1 - \frac{\mathrm{sgn}(E) \, \Gamma^{\mathrm{diff}}}
  {\max(|\Gamma^{\mathrm{drift}}|, \epsilon)},
\end{equation}
where all quantities are defined at the cell face, $\mathrm{sgn}(x)$ is the sign
function, and $\epsilon$ is a small number to avoid division by zero. In 2D or
3D, the term $\mathrm{sgn}(E)\,\Gamma^{\mathrm{diff}}$ would be replaced by
$\hat{E}_j\,\Gamma_j^{\mathrm{diff}}$, where $\hat{E}_j$ is the
$j$\textsuperscript{th} component of the electric field unit vector, and
$|\Gamma^{\mathrm{drift}}|$ would be replaced by $|\Gamma_j^{\mathrm{drift}}|$.

When $f_\epsilon$ is computed at cell faces, it makes sense to also evaluate the
source term at cell faces, so that equation (\ref{eq:source-gamma}) becomes
\begin{equation}
  \label{eq:soloviev-flux}
  S_i = \left(\bar{\alpha}_L f_{\epsilon,L} |\Gamma_{L}^\mathrm{drift}| +
    \bar{\alpha}_R f_{\epsilon,R} |\Gamma_{R}^\mathrm{drift}|\right)/2,
\end{equation}
where the subscripts $L$ and $R$ indicate values on the left and right face of
the cell. In this paper, the $f_\epsilon$ scheme is implemented according to
equation (\ref{eq:soloviev-flux}). Alternatively, it is also possible to
approximate $f_{\epsilon}$ at the cell center, for example as
\begin{equation*}
  f_\epsilon = 1 - \frac{\mathrm{sgn}(E_L) \Gamma_L^{\mathrm{diff}} +
    \mathrm{sgn}(E_R) \Gamma_R^{\mathrm{diff}}}
  {|\Gamma_L^{\mathrm{drift}}| + |\Gamma_R^{\mathrm{drift}}| + \epsilon},
\end{equation*}
after which the standard source term of equation (\ref{eq:source-cc}) can be
used.

Note that $f_\epsilon$ can in principle become negative, which would be
unphysical. Here $f_\epsilon$ is set to zero in such cases. Similarly,
$f_\epsilon$ can be restricted to be at most one, to prevent an increase in the
source term.

\subsection{Source term from flux (FFS scheme)}
The source term in equation (\ref{eq:source}) can also be approximated by the
full electron flux
\begin{equation}
  \label{eq:src-flux}
  S = \bar{\alpha} \mu_e |E| n_e = \bar{\alpha} |\Gamma^\mathrm{drift}| \approx \bar{\alpha} |\Gamma|,
\end{equation}
where $\Gamma$ is given by equation (\ref{eq:flux-drift-diffusion}). This
approach is here referred to as the FFS (Full Flux Source) scheme. The
FFS scheme has a similar effect as the factor introduced in equation
(\ref{eq:soloviev}). When the advective and diffusive fluxes balance each other,
the source term is zero. The effective factor introduced in equation
(\ref{eq:src-flux}) is $g = |\Gamma|/|\Gamma^\mathrm{drift}|$. When the advective
and diffusive flux are (anti-)parallel, which is always the case in 1D, it
follows that $g = f_\epsilon$. For example, when the fluxes are in opposite directions
$g = 1 - |\Gamma^{\mathrm{diff}}|/|\Gamma^{\mathrm{drift}}|$, just as for equation
(\ref{eq:soloviev}). In general 2D or 3D cases, $g \neq f_\epsilon$. When the advective
and diffusive flux are orthogonal,
$g = \sqrt{1 + (|\Gamma^{\mathrm{diff}}|/|\Gamma^{\mathrm{drift}}|)^2}$, whereas
$f_\epsilon = 1$. However, in such cases the diffusive flux is typically small compared
to the advective flux, so that
$g \approx 1 + (|\Gamma^{\mathrm{diff}}|/|\Gamma^{\mathrm{drift}}|)^2/2 \approx 1$.

Compared to equation (\ref{eq:soloviev}), there is less physical motivation for
equation (\ref{eq:src-flux}). Both approaches have the drawback that the
original model formulation is changed. However, numerically equation
(\ref{eq:src-flux}) is easier to implement than equation (\ref{eq:soloviev}),
since the electron flux is readily available. The source term can be computed
similar to equation (\ref{eq:source-gamma}), so that in 1D it is given by
\begin{equation}
  \label{eq:g-scheme}
  S = \left(\bar{\alpha}_L |\Gamma_L| + \bar{\alpha}_R |\Gamma_R|\right)/2,
\end{equation}
where $\Gamma_L$ and $\Gamma_R$ are the flux through the left and right face of
the cell. Finally, note that equation (\ref{eq:src-flux}) can be modified so
that it does not increase the source term by replacing $|\Gamma|$ with
$\min(|\Gamma^\mathrm{drift}|, |\Gamma|)$.

\subsection{Limiting parallel diffusion}
\label{sec:limit-parall-diff}

With the above two methods, the source term is modified next to a strong density
gradient. This prevents electrons that have diffused across the gradient from
generating new ionization. Another approach could be to limit diffusion parallel
to the electric field, which can be implemented as:
\begin{equation*}
  \textnormal{if }\Gamma_j^\mathrm{diff} E_j > 0 \textnormal{ and }
  |E_j| > E_\mathrm{lim} \textnormal{ then set } \Gamma_j^\mathrm{diff} = 0.
\end{equation*}
Or in words: if the $j$\textsuperscript{th} component of the diffusive flux
points in the same direction as $E_j$, and if $|E_j|$ exceeds some threshold
$E_\mathrm{lim}$, set the diffusive flux to zero. By setting $E_\mathrm{lim}$ to
the critical field, above which the net ionization rate is positive, electrons
cannot diffuse into a region where they generate ionization.

Advantages of this approach are that only the diffusion equation has to be
modified and that the implementation is relatively simple. However, there are
also some downsides. As long as the electron density is not strictly zero,
ionization continues to occur when the field is above breakdown. For the
implementation described above, it is also necessary to evaluate the source term
using the electron drift flux, as in equation (\ref{eq:source-gamma}).
Otherwise, electrons can diffuse in from one of the cell faces (with a low
field), and generate ionization at the cell center due to a high field on the
other cell face.

% \begin{figure}
%   \centering
%   \includegraphics[width=0.5\textwidth]{fig_diffusion_Dprob.pdf}
%   \caption{When diffusion parallel to the field is limited, the density in a
%     single cell can continue to grow. The dashed line shows the cell-centered
%     electric field, which is the average of the fields on the cell faces. Where
%     the peak forms, the cell-centered field stays above breakdown, as explained
%     in the text.}
%   \label{fig:diffusion-dprob}
% \end{figure}

\subsection{Comparison of the schemes}
\label{sec:tests-diff-schemes}

\begin{figure}
  \centering
  \includegraphics[width=0.5\textwidth]{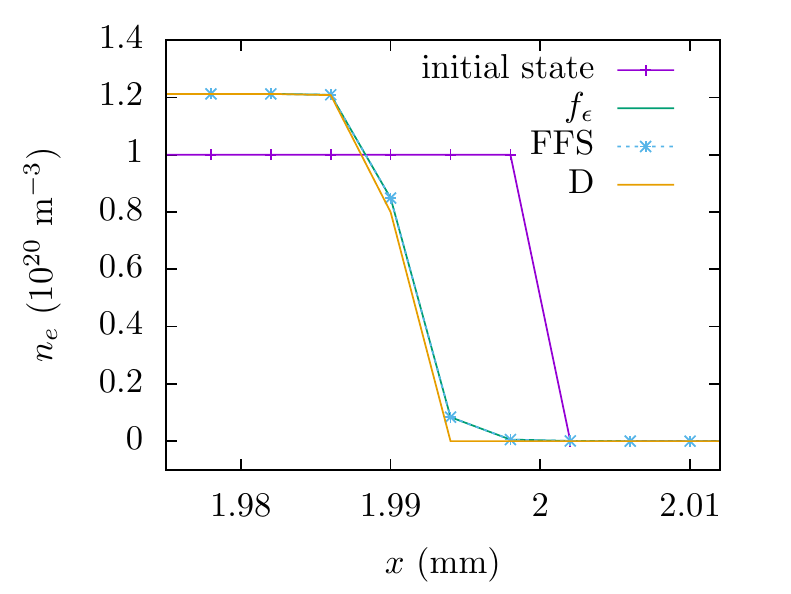}
  \caption{Comparison of methods to address unphysical diffusion, showing the
    electron density at $t = 8 \, \textrm{ns}$ for the test case described in
    section \ref{sec:diff-test-case}. The $f_\epsilon$ curve corresponds to
    equation (\ref{eq:soloviev}), the $FFS$ curve to equation
    (\ref{eq:src-flux}) and the $D$ curve to the diffusion-limited approach
    described in section \ref{sec:limit-parall-diff}. The $f_\epsilon$ and FFS
    scheme give identical results. All methods prevent unphysical growth of the
    discharge to the right.}
  \label{fig:diff-fix-1}
\end{figure}

In figure \ref{fig:diff-fix-1}, the effect of the three approaches described
above is compared. Applied to the half-ionized domain test case described in
section \ref{sec:diff-test-case}, the results at $t = 8 \, \textrm{ns}$ show no
unphysical propagation to the right. As expected, equation (\ref{eq:soloviev})
and equation (\ref{eq:src-flux}) give identical results in 1D. With the
diffusion-limited approach, the electron density drops to zero more rapidly.
From this 1D test, it cannot be concluded which scheme will perform best for
general 2D and 3D cases; this question is left for future work.

\begin{figure}
  \centering
  \includegraphics[width=0.5\textwidth]{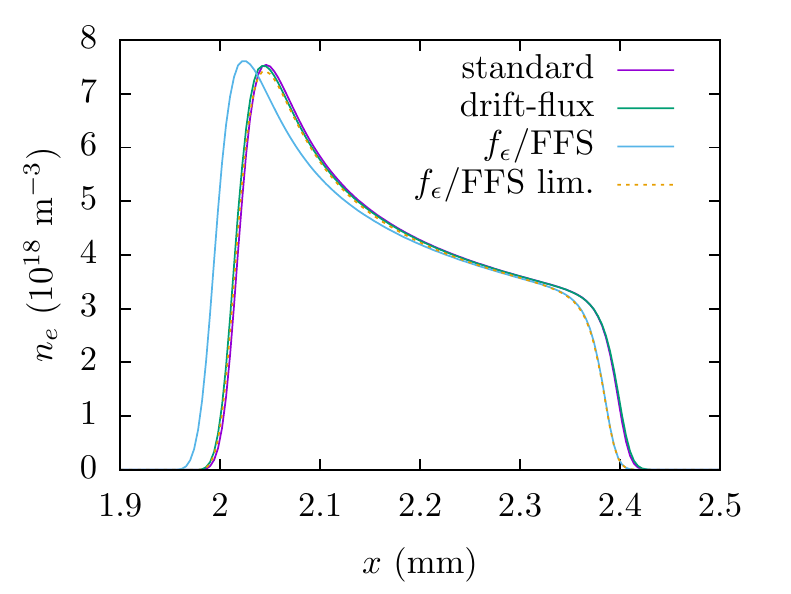}
  \caption{Comparison of several methods to compute the ionization source term
    for the test case described in section \ref{sec:tests-diff-schemes}. The
    electron density at $t = 3 \, \textrm{ns}$ is shown. Using equation
    (\ref{eq:source-cc}) for the source term (labeled \emph{standard}) produces
    almost the same results as equation (\ref{eq:source-gamma}) (labeled
    \emph{drift-flux}). The $f_\epsilon$/FFS schemes can increase the source
    term, which causes the discharge to propagate slightly faster. When the
    source is limited (so that it does not exceed its `normal' value) this
    effect disappears, as indicated by the dashed line.}
  \label{fig:diff-fix-2}
\end{figure}

In this paper, several approaches have been described for computing the
ionization source term. Their effect on a developing discharge is illustrated in
figure \ref{fig:diff-fix-2}, which shows results for four cases:
\begin{itemize}
  \item A standard DD-LFA model, using equation (\ref{eq:source-cc}) as the source term
  \item A DD-LFA model that uses the drift flux for the source term, see equation (\ref{eq:source-gamma})
  \item The $\epsilon$ and FFS schemes (which give identical results)
  \item The $\epsilon$ and FFS schemes, modified such that they do not increase the ionization rate
\end{itemize}
The same computational domain and gas as in section \ref{sec:diff-test-case} are
used, but with $30 \, \textrm{kV}$ applied voltage. Initially,
$n_e = n_p = 10^{10} \, \mathrm{m}^{-3}$ for
$2.95 \, \mathrm{mm} \leq x \leq 3.05 \, \mathrm{mm}$, and zero elsewhere. All
simulations were performed with a fixed time step $\Delta t = 3 \, \mathrm{ps}$.

With the $\epsilon$/FFS schemes, the source term on the right side of the
discharge is reduced, leading to a smaller extension in this direction.
Conversely, these schemes increase the source term on the left side, leading to
faster leftward propagation than with the standard model. Having a higher
ionization rate in front of the discharge physically makes sense~\cite{Li_2007},
but the changes introduced by equations (\ref{eq:soloviev}) and
(\ref{eq:src-flux}) are not necessarily the best way to correct for this. When
the $\epsilon$/FFS schemes are modified so that they do not increase the source
term, the results are in good agreement with the standard model for the leftward
propagation. Finally, figure \ref{fig:diff-fix-2} shows that equation
(\ref{eq:source-cc}) and equation (\ref{eq:source-gamma}) give almost the same
results for this test case.

\section{Summary and discussion}
\label{sec:conclusions}

The main contributions of the present paper are:
\begin{itemize}
  \item A new explicit approach to avoid the dielectric relaxation time step
  restriction was presented in section \ref{sec:diel-relax-time}. Compared with
  the existing semi-implicit method, the new current-limited approach is faster,
  simpler to implement, does not reduce the order of accuracy of the time
  integrator, and it reverts to the original model for sufficiently small time
  steps. The proposed current-limited approach can be used with different types
  of plasma fluid models.
  \item With the local field approximation, unphysical effects can occur when
  there is diffusion parallel to the electric field. Several methods to avoid
  this were compared in section \ref{sec:diffusion-grad}. An existing
  approach~\cite{Soloviev_2009} was compared with two new approaches. All
  methods could prevent the unphysical effects in 1D.
  \item Different methods for implementing the ionization source term were
  described, as well as implementation details for the approaches listed above.
  General implementation aspects for explicit drift-diffusion fluid models were
  also discussed.
\end{itemize}

The methods described here are particularly relevant for the simulation of
pulsed discharges near dielectrics and electrodes. Near such boundaries, there
can be high electron densities, strong electric fields, and strong spatial
gradients in both quantities, see e.g.~\cite{Babaeva_2016}. For simplicity,
tests were here performed in 1D, although the described methods can also be used
in 2D and 3D. Performing these 2D and/or 3D tests is an important next step that
is left for future work. For the unphysical effects due to diffusion, it would
also be interesting to investigate the effect of using an energy equation
instead of the local field approximation.

\section*{Acknowledgments}

JT was supported by fellowship 12Q6117N from Research Foundation -- Flanders
(FWO) and by the State Key Laboratory of Electrical Insulation and Power
Equipment (EIPE18203). Robert Marskar is acknowledged for pointing out reference
\cite{Soloviev_2009}.

\section*{Software availability}

The source code of the simulation software used in this paper is available at:
\url{https://gitlab.com/MD-CWI-NL/streamer_1d} and
\url{https://github.com/jannisteunissen/semi_implicit_fluid}.

\section*{References}
\bibliographystyle{iopart-num}
\bibliography{references.bib}

% \subsection{Local field approximation}

% Since $\sigma = e \mu n_e$, we have to be extra careful when using the so-called
% \emph{local field approximation}, in which $\mu$ is a function of the electric
% field strength $E$. For $E \rightarrow 0$ the mobility often has a sharp peak,
% because electrons lose less momentum (and energy) over time at lower energies.

% Where $n_e$ is large, electric fields will be small, and the mobility can be
% quite high with the local field approximation. I believe this is unrealistic,
% because the lack of loss mechanisms means electrons lose energy slowly when they
% enter such a region. Limiting the mobility to a reasonable value can thus
% potentially make simulations \emph{more} realistic.

% \subsection{Examples}

% TODO!

\end{document}